\documentclass[a4paper,11pt]{article}
\pdfoutput=1 

\def\eV{\si{\electronvolt}\xspace}
\def\um{\si{\micro\m}\xspace}

\def\dedx{\ensuremath{\mathrm{d}E/\mathrm{d}x}\xspace}

\def\Fe{\ensuremath{^{55}}Fe\xspace}

\def\Ka{K\ensuremath{_{\upalpha}}\xspace}
\def\Kb{K$_{\upbeta}$\xspace}
\def\kapton{Kapton\texttrademark\xspace}

\def\co{CO$_2$\xspace}
\def\n{N$_2$\xspace}

\newbox\mytempbox
\newdimen\mytempdimen

\def\VLP{$\Delta\!V_\mathrm{LP}$\xspace}
\def\VS{$\Delta\!V_\mathrm{S}$\xspace}
\def\VSP{$\Delta\!V_\mathrm{SP}$\xspace}

\def\Ecoll{$E_\mathrm{coll}$\xspace}
\def\Eextr{$E_\mathrm{extr}$\xspace}

\usepackage[utf8]{inputenc}
\usepackage{siunitx}
\usepackage{amsmath}
\usepackage{amssymb}
\usepackage{amsthm}
\usepackage{upgreek}
\usepackage{eurosym}
\usepackage{xspace}
\usepackage{textpos}
\usepackage[obeyDraft]{todonotes}
\usepackage{lineno}

\usepackage{jinstpub} 

\title{Ion backflow studies with a triple-GEM stack with increasing hole pitch}

\author[a,1]{H.~Natal~da~Luz\note{Corresponding author.}}
\author[b]{P.~Bhattacharya}
\author[a]{L.A.S.~Filho}
\author[a]{L.E.F.M.~Fran\c{c}a}

\affiliation[a]{Instituto de F\'isica da Universidade de S\~ao Paulo\\Rua do Mat\~ao 1371, 05508-090 Cidade Universit\'aria, S\~ao Paulo, Brasil}
\affiliation[b]{Department of Particle Physics and AstroPhysics, Weizmann Institute of Science, 234 Herzl St., Rehovot, Israel}

\emailAdd{hugonluz@if.usp.br}

\abstract{Gas Electron Multipliers have undergone a very consistent development since their invention in 1997. Their production procedures have been tuned in such a way that nowadays it is possible to produce foils with areas of the order of the square meter that can operate at a reasonable gain, uniform over large areas and with a good stability in what concerns electrical discharges. For the 3$^\mathrm{rd}$ run of LHC, they will be included in the CMS and ALICE experiments after significant upgrades of the detectors, confirming that these structures are suitable for very large experiments.
In the special case of Time Projection Chambers, the ion backflow and the energy resolution are sensitive issues that must be addressed and the GEM has shown to be able to deal with both of them. 

In this work, a stack of three GEMs with different pitches has been studied as a possible future approach for ion backflow suppression to be used in TPCs and other detection concepts. With this approach, an ion backflow of 1\,\% with an energy resolution of 12\,\% at 5.9\,k\eV has been achieved with the detector operating in an Ar/\co (90/10) mixture at a gain of $\sim 2000$.}


\arxivnumber{1803.09382} 

\begin{document}
\maketitle
\flushbottom

\section{Introduction}
\label{sec:intro}

Gas Electron Multipliers (GEM)~\cite{Sau97,Sau16} and other Micropattern Gaseous Detectors, such as the MICROMEGAS~\cite{Gio96} and Thick-GEMs~\cite{Bre09} became established technologies for very large detectors and their use is foreseen for upgrades in some High Energy Physics experiments, such as ALICE~\cite{ALI13}, CMS~\cite{CMS15} or COMPASS~\cite{COM16}. The choice for this type of detectors is related to their cost per unit of sensitive area, moderate energy and position resolution and other less obvious -- although important -- effects, such as their intrinsic ion backflow (IB) suppression. IB is the flow of the positive ions resulting from the electron avalanches in the gas in the direction of the drift region. The presence of these ions in the drift region of the detector gives origin to distortions in the electric field that in most of the cases should be uniform. Not every detector concept is sensitive to IB but, for example, in Time Projection Chambers (TPC) the reconstruction of particle tracks relies on the assumption that the drift field is uniform to estimate the $x,y,z$ position of each voxel. Depending on the amount of IB, the position and specific energy loss (\dedx) resolution can be compromised.

There have been consistent and successful studies for the optimization of GEM geometries and high-voltage parameters to reduce the IB enough to make these structures suitable for their use in a continuously operated TPC, which is a technological breakthrough, increasing its performance in terms of event rate capability~\cite{Gas14,Sai15,Dei16}. In case of the ALICE TPC upgrade the geometry studies converged to a cascade of four GEMs where the first and the last ones use a standard 140\,\um hole pitch and the two in between have a hole pitch of 280\,\um. The experiment requires IB below 1\,\% and energy resolution below 12\,\% at 5.9\,k\eV ($\sigma/E$). The chosen geometry fulfilled the requirements with a gas mixture of Ne/\co/\n (90/10/5) and specific HV settings. The large amount of degrees of freedom in the high voltage parameters gives a very good flexibility for this setup, specially due to the fact that the operational conditions that optimize the IB without exceedingly compromising the energy resolution demands voltage configurations that are usually less stable in terms of discharge probability.

Despite the advantages of the quadruple GEM, it is interesting to address and compare the performance of a simpler, triple GEM stack. Therefore, we tried to benefit from the collection and extraction efficiencies of GEMs with different hole pitches to optimize the IB suppression of this stack. The work described in this paper studied the performance of a triple GEM stack where each of the three GEMs have a different pitch, operating in an Ar/\co (90/10) mixture. The work consisted of laboratory measurements and numerical simulations using Garfield~\cite{Vee93}.

\section{Experimental setup}
\label{sec:experimental}

Throughout this text, the detector is described with the cathode on top with the electrons moving down towards the anode. The nomenclature of the fields follow the standard used in this type of detectors. The field above the stack, where the radiation is absorbed and the primary eletrons drift towards the first GEM is the drift field, and the field below the stack where the electrons are pulled towards the anode is the induction field. The field between two GEMs is the transfer field and can be called the extraction field (\Eextr) of the GEM at the top or the collection field (\Ecoll) of the GEM at the bottom. The GEMs in a stack are numbered from top to bottom.

\subsection{Basic principle}
\label{subsec:basic}

The working principle proposed for this GEM stack is based on two important concepts related to the GEM: the collection efficiency and the extraction efficiency. 

The collection efficiency is defined as the fraction of free electrons drifting towards the GEM that enter the holes. The ability of the GEM to focus the electrons towards the holes depends on the ratio between the drift field (the field above the GEM) and the field inside the holes. 
At moderate drift fields, just enough to prevent the recombination of the ion pairs, the drift electric field lines are mostly focused to the holes of the GEM. In this setup, most of the electrons in the drift region are able to reach the holes and the collection efficiency is close to unity. If the drift electric field is increased with respect to the GEM voltage, the field lines will curve towards the GEM holes closer to the GEM copper surface. Above a certain threshold, some of the field lines will no longer curve towards the holes, ending on the copper surface between them. As a consequence, some of the electrons in the drift region are lost to the copper surface, resulting in a continuous reduction of the collection efficiency as the drift field further increases~\cite{Bac99}. To restore the collection efficiency, the voltage across the GEM must be increased (with the consequent increase of its absolute gain)~\cite{Sau06}. It is relatively simple to understand that if the hole pitch is larger, the threshold in the drift field for collection efficiency reduction will be at a lower drift field~\cite{Bac99} due to the smaller ratio between the area of the holes and the whole area of the GEM (defined here as the optical transparency), and the opposite happens for smaller pitch. Ref.~\cite{Luz18} shows a comparative study of the anode current as the drift field changes for GEMs and Thick-GEMs of different hole pitch. 

After being multiplied, the electrons must be pulled out of the holes. Some of them will be collected on the bottom copper surface of the GEM and some of them will make their way through the induction field towards the anode. The extraction efficiency is defined as the fraction of electrons leaving the GEM and reaching the anode plane. The extraction depends on the ratio between the induction field (below the GEM) and the field inside the holes. When the voltage across the GEM is kept constant, at low induction fields, most of the electric field lines leaving the holes are curved back to the bottom copper surface of the GEM, resulting in a very small number of electrons reaching the anode. An increase of the induction field will bring these lines closer to the anode before they bend and more electrons reach the anode, increasing the extraction efficiency more or less linearly. Again, above a certain threshold, the induction field is so high that there are no more lines bending from the holes towards the bottom surface of the GEM. In this regime, the current in the anode continues to increase at a smaller rate, partly due to the increase of the extraction efficiency and partly due to the influence of the induction field in the field inside the GEM holes, which increases its absolute gain~\cite{Bac99}. Eventually, at some even higher field, the current increases again at a higher rate due to the onset of Townsend avalanches also in this region. The threshold will also move for lower induction fields when the hole pitch is larger and the opposite will happen for smaller pitch.

%
%
%
%
%

For the case of ions, the same effect takes place, but coming from below the GEM in the opposite direction. Taking that into account, the way to increase the stream of electrons extracted from one GEM while reducing the number of ions in the opposite direction is to increase the induction field. In fact, as seen before, the electron extraction efficiency increases 
while the amount of ions entering the holes from below decreases. 
However, in a GEM stack, although the transfer field is the extraction field of the GEM on the top, it is the collection field of the GEM at the bottom, which means that the advantages of increasing this field to improve the extraction from the first GEM will be lost by the loss of collection efficiency on the second GEM. 
The results relating the electron transparency with the optical transparency~\cite{Bac99} suggest that using a second GEM with a smaller pitch, the field can go higher while keeping the collection efficiency close to unity. 

A very good model describing these effects and their consequences, including parametrization of both the collection and extraction efficiencies can be found in~\cite{Kil03}. The ratio between both is described in terms of the external field and the field inside the holes of the GEMs:

\begin{equation}
\label{eq:ratio}
\frac{X}{C} = \frac{1}{T_\mathrm{opt}}\frac{E_\mathrm{ext}}{E_\mathrm{h}},
\end{equation}
where   $T_\mathrm{opt}$ is the optical transparency of the GEM, defined here as the ratio between the area of the holes and the total area of the GEM, $E_\mathrm{ext}$ and $E_\mathrm{h}$ are the external field and the field inside the holes of the GEM. $X$ and $C$ are the extraction and collection efficiencies, for ions/electrons and electrons/ions, according to the side to which $E_\mathrm{ext}$ refers to. The extraction and collection efficiencies can be represented as $X^+$ and $C^+$ for ions and $X^-$ and $C^-$ for electrons. For example, the effective gain of a single GEM detector is given by:

\begin{equation}
\label{eq:geff}
G_\mathrm{eff} = C^- G_\mathrm{abs} X^-,
\end{equation}

\noindent where $G_\mathrm{eff}$ and $G_\mathrm{abs}$ are the effective and absolute gain, respectively.
 By definition, the IB is the number of positive ions reaching the cathode ($N_\mathrm{i}$) per number of electrons reaching the anode ($N_e$). For a triple GEM stack, this can be expressed in terms of the extraction and collection efficiencies for ions and electrons ($X_i^\pm$ and $C_i^\pm$) and the absolute gain ($G_{i}$) of each GEM:

\begin{equation}
\label{eq:ib1}
IB = \frac{N_i}{N_e} = \frac{C_1^+ G_{1} X_1^+\cdot C_2^+ G_{2} X_2^+\cdot G_{3} X_3^+ + N_\mathrm{p}}{N_\mathrm{p} \cdot C_1^- G_{1} X_1^- \cdot C_2^- G_{2} X_2^- \cdot C_3^- G_{3} X_3^-},
\end{equation}
\noindent where $C_1^- G_{1} X_1^-\cdot C_2^- G_{2} X_2^-\cdot C_3^- G_{3} X_3^-$, from the denominator, is the effective gain $G_\mathrm{eff}$ of a triple GEM stack and $N_\mathrm{p}$ is the number of primary electron-ion pairs.

Reorganizing and replacing according to equation \ref{eq:ratio}, naming the external fields following the convention of GEM detectors, one gets:

\begin{equation}
\label{eq:ib2}
\begin{array}{r c l}
IB  & = &  \frac{X_1^+}{C_1^-} \cdot \frac{C_1^+}{X_1^-} \cdot \frac{X_2^+}{C_2^-} \cdot \frac{C_2^+}{X_2^-} \cdot \frac{X_3^+}{C_3^-} \cdot \frac{1}{N_\mathrm{p} X_3^-}  \cdot + \frac{1}{G_\mathrm{eff}} \\
 & & \\
 & = & \frac{1}{T_{\mathrm{opt}1}}\frac{E_\mathrm{drift}}{E_\mathrm{h1}} \cdot  T_{\mathrm{opt}1}\frac{E_\mathrm{h1}}{E_\mathrm{trans1}} \cdot \frac{1}{T_{\mathrm{opt}2}}\frac{E_\mathrm{trans1}}{E_\mathrm{h2}} \cdot  T_{\mathrm{opt}2}\frac{E_\mathrm{h2}}{E_\mathrm{trans2}} \cdot \frac{1}{T_{\mathrm{opt}3}}\frac{E_\mathrm{trans2}}{E_\mathrm{h3}} \cdot \frac{1}{N_\mathrm{p} X_3^-}  \cdot + \frac{1}{G_\mathrm{eff}},
\end{array}
\end{equation}
where $E_\mathrm{drift}$ is the field in the drift region, $E_{\mathrm{h}i}$ is the field inside the holes of GEM $i$,  $E_{\mathrm{trans}i}$ is the transfer field below GEM $i$ and $T_{\mathrm{opt}i}$ is the optical transparency of GEM $i$. Simplifying the whole expression results in:

\begin{equation}
\label{eq:IB}
IB = \frac{E_\mathrm{drift}}{T_{\mathrm{opt}3} E_\mathrm{h3} N_\mathrm{p} X_e^-} + \frac{1}{G_\mathrm{eff}}.
\end{equation}

This expression confirms the well known effects of how the IB increases by increasing the magnitude of the drift field or by decreasing the gain in GEM-like detectors. It also shows that for the same effective gain of the detector, increasing the absolute gain of the last GEM will help reducing the IB, which is somehow expected, since this means the absolute gain of the GEMs closer to the cathode must be reduced, reducing the number of ions in that region. 

It is also very interesting how the optical transparency of the last GEM can also have a great influence in blocking the ions. This is the main principle exploited in this work. In fact, by using three GEMs with increasing optical transparency, it is possible to apply transfer fields that provide a high electron extraction efficiency from one GEM, while still providing a high electron collection efficiency in the following one.

\subsection{The detector}
\label{subsec:detector}

\begin{figure}[htbp]
\begin{center}
\includegraphics[width=0.7\textwidth]{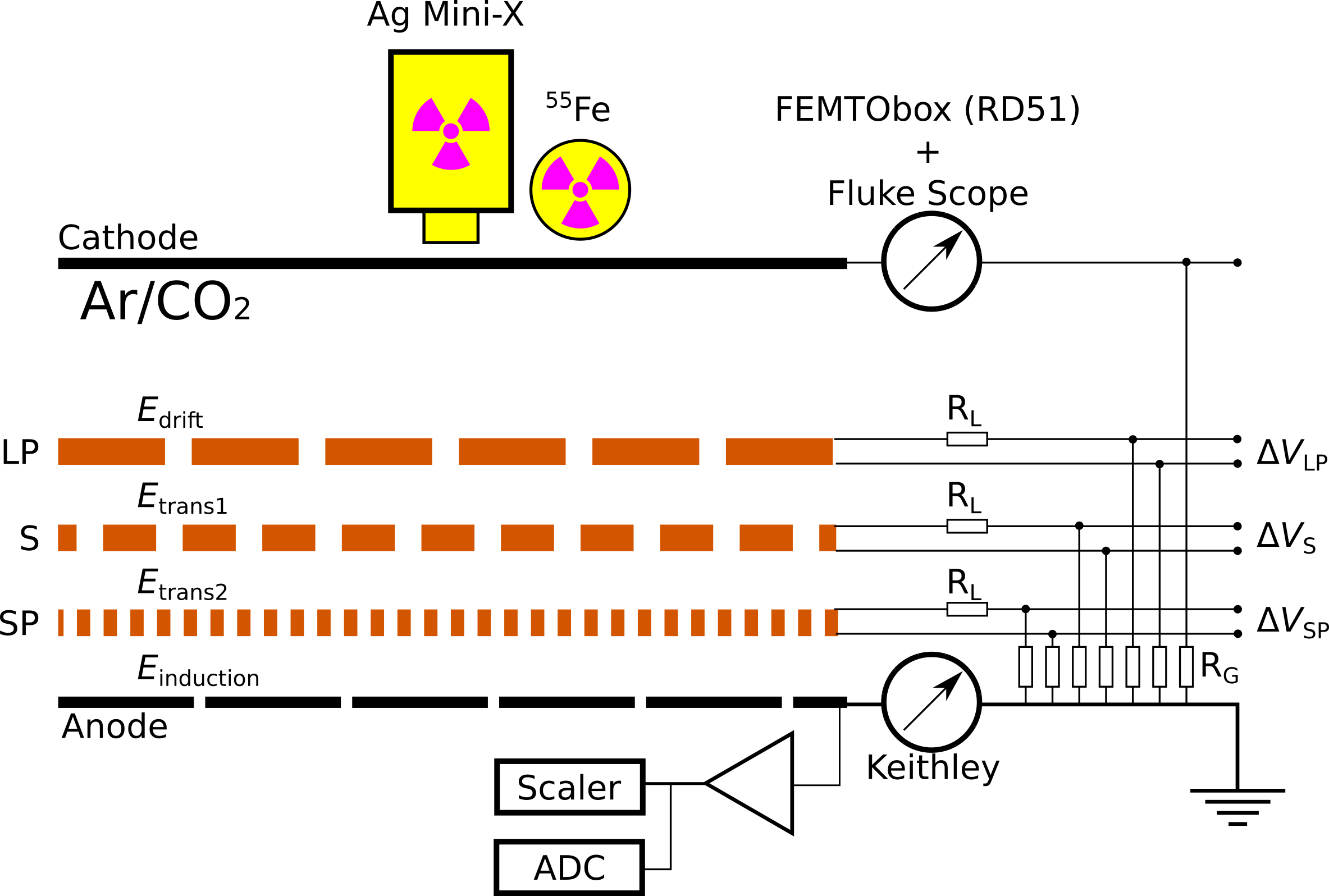}
\caption{Experimental setup. The GEM stack is composed of three GEMs with pitch 280\,\um, 140\,\um and 90\,\um. 
The gain and the energy resolution were measured using a \Fe radioactive source.}
\label{fig:setup}
\end{center}
\end{figure}

The experimental setup used for this study is shown in fig.~\ref{fig:setup}. A triple GEM stack was mounted using 50\,\um thick GEMs built at CERN with hole pitch 280\,\um (LP, top GEM), 140\,\um (S, middle GEM) and 90\,\um (SP, bottom GEM). The cathode was made of a copper clad 50\,\um \kapton foil. The drift, transfer 1 and 2, and induction gaps were 7.2, 2.2 and 1.6\,mm thick, respectively. The readout was composed of 120 $10\times 8.2\,\si{\square\milli\m}$ pads covering the whole $10\times 10\,\si{\square\cm}$ area of the GEMs. 

An Ar/\co (90/10) gas mixture was used in open flow at a rate around 6\,l/h. To generate the electrons inside the drift gap, either an \Fe radioactive source or an Amptek Mini-X silver target X-ray tube were used. 

The detector was biased using seven independent high voltage channels from CAEN VME power supplies, through loading resistors connected to the top electrode of each GEM and grounding resistors that protected the GEMs in case of electrical discharge, draining the excess current to the ground instead of forcing it into the HV supply.

The current in the anode was measured with a Keithley 6517b electrometer and the current in the cathode was measured with a floating femtoamperemeter developed at CERN within the RD51 collaboration. In the femtoamperemeter, an operational amplifier measures the voltage at a silicon diode, allowing to measure currents over a very wide range. It was calibrated by plotting the currents measured by the electrometer against the voltage supplied by the diode, read by a Fluke Scopemeter 124 handheld oscilloscope. To supply the currents, the cathode was biased with a positive voltage while ions were collected on the top electrode of the first GEM, when the gas was irradiated by X-rays. The charge pre-amplifier, shaping amplifier, logic and ADC for the pulse height distributions were all standard nuclear instrumentation NIM and CAMAC modules. The pulse height distributions and the current measurements were performed simultaneously by reading the current induced through a coupling capacitor. The measurements of the current were tested also without the capacitor and the values were the same, excluding the possibility of charge sharing. The IB and the energy resolution measurements were done simply by interchanging the two X-ray sources, keeping the voltage settings. 

It is known that high ion space-charge charge density in the drift region near the GEM influence systematically the IB~\cite{Bal14}. This charge density affects mainly the ions. The currents measured in the cathode $I_\mathrm{drift}$ were around 100\,pA. Since the X-ray source was collimated by a 2\,mm diameter hole and the drift gap had a length of 7.2\,mm, the areal charge density $\sigma_\mathrm{Q}$ (given by the product of the charge density $\rho$ by the drift length $d$) is given by~\cite{Bal14}:
\begin{equation}
\sigma_\mathrm{Q} = \frac{I_\mathrm{drift} \cdot d}{A\cdot v_\mathrm{ion}},
\end{equation}
\noindent where $A$ is the area covered by the X-ray beam and $v_\mathrm{ion}$ is the ion drift velocity for the electric field and the gas mixture used. Assuming a drift velocity of 450\,\si{\cm\per\second} for argon ions in Ar-CO$_2$ with a field of 300\,V/cm, the areal charge density is around 5\,pC/cm$^2$, below the $10^4$\,fC/cm$^2$ measured as the limit above which there is a strong decrease in the measured IB. The typical counting rate was around 1\,kHz/cm$^2$, well under the limit for which the high charge densities start affecting the gain in a GEM~\cite{Res15}.


The gain, tuned to $2000\pm 50$ throughout all the scans, was measured by counting the rate of the \Ka peak of the manganese line from the \Fe radioactive source to obtain the primary current and correcting the anode current to remove the small contribution of the argon escape peak. This gain was used as a reference in order to have a comparison term with the baseline settings of ALICE, which is the most systematic study carried out so far, with all parameters unambiguously described~\cite{ALI13}.

\section{Results and discussion}
\label{sec:results}

\begin{figure}[htbp]
\begin{center}
\includegraphics[width=0.6\textwidth]{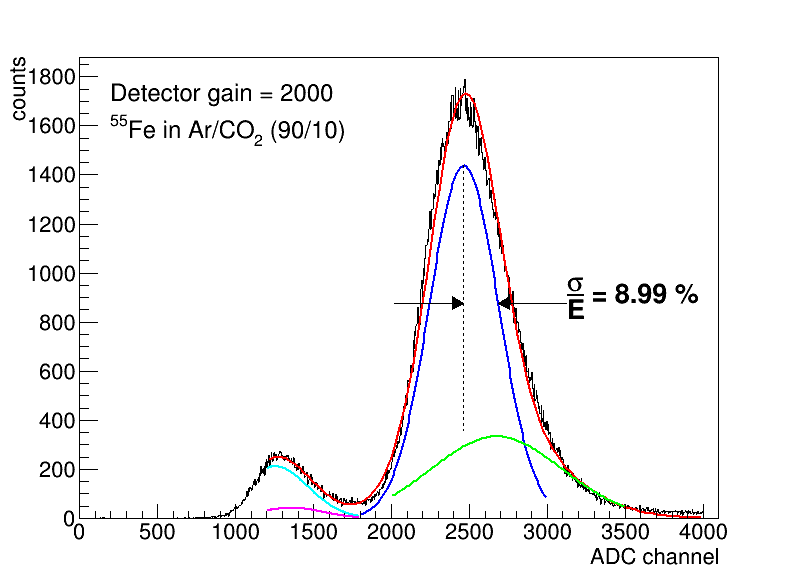}
\caption{Pulse height distribution from an \Fe source, with the detector operating at a charge gain 2000. This distribution was obtained with a non collimated source placed on the detector \kapton window ($\sim$1\,cm from the cathode). It shows a very good resolution for this gain regime, demonstrating the good quality of the GEMs manufactured at CERN. The main peak was fit with two gaussians, corresponding to the \Ka and \Kb at 5.9 and 6.5\,keV, respectively. The smaller peak is the argon escape peak and was fitted with two gaussians centered in the K-lines minus the argon absorption edge. The voltages are not optimized for IB suppression. }
\label{fig:espectro}
\end{center}
\end{figure}

Figure~\ref{fig:espectro} shows a pulse height distribution obtained with a \Fe radioactive source, with the detector operating at gain 2000. Although the energy resolution ($\sigma_\mathrm{E}$), defined as the ratio between the standard deviation and the centroid of a gaussian distribution fitted to the pulse height distribution, does not depend only on the hole size distribution uniformity, the reliability of the GEMs manufactured at CERN nowadays is well known and the areal non-uniformity for small foils ($10\times 10\,\si{\square\cm}$) such as the ones used in this work does not seam to play an important role. In fact, even without collimating the X-ray source, it was possible to achieve a very satisfactory resolution. The settings used for this pulse height distribution were not optimized for ion backflow suppression.

\subsection{Experimental results}
\label{subsec:expresults}



\begin{figure}[htbp]
\begin{center}
\includegraphics[width=0.6\textwidth]{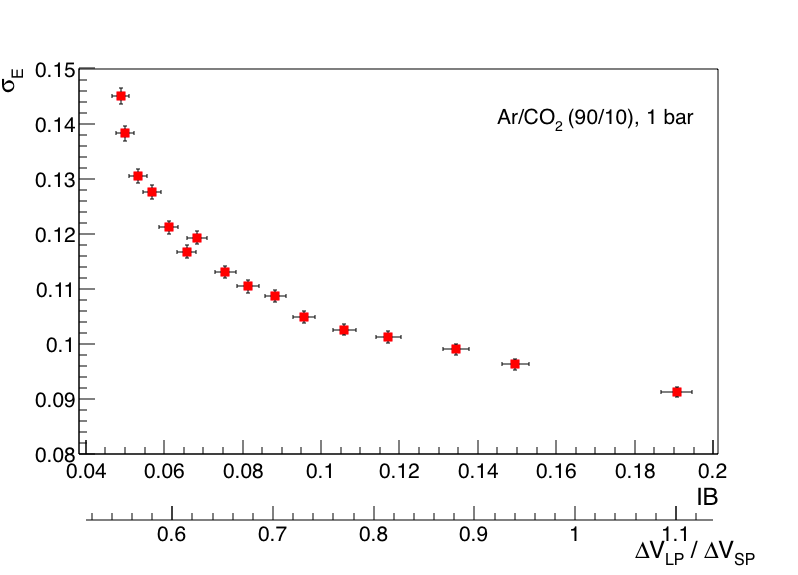}
\caption{$\Delta\!V_\mathrm{LP}$ and $\Delta\!V_\mathrm{SP}$ scan. Since the gain was kept constant, when \VLP increased, \VSP decreased and vice-versa. The ratio \VLP/\VSP is shown as the secondary $x$-axis to show that the ion backflow increases by decreasing \VSP. The ion backflow and the energy resolutions are two competing properties and, in this scan, whenever one of them improves, the other worsens. This is not necessarily true for all the parameters.}
\label{fig:VlpVsp}
\end{center}
\end{figure}

The energy resolution and ion backflow are two properties that are competing against each other. 
Figure~\ref{fig:VlpVsp} depicts the energy resolution for IB values obtained with different high voltage settings in this experimental setup. The data points were measured by scanning the voltage of the first GEM in the stack (the LP GEM), while tuning the gain by changing the voltage of the last GEM. None of the other voltages was optimized for IB suppression, but it clearly shows that decreasing the voltage of the first GEM to reduce the IB has the effect of increasing the energy resolution and vice versa. The decrease of the IB as the voltage in the last GEM increases is understood from eq.~\ref{eq:IB}. In fact, by increasing the voltage in the last multiplication stage while decreasing it in the first GEM the IB is reduced due to the smaller number of ions in its holes  --- they are mostly concentrated in the last GEM and must drift through the first two GEMs in a stack before reaching the drift region. A drawback is the deterioration of the energy resolution in optimal IB conditions.

In GEM detectors, each multiplication or transfer stage is introducing statistical fluctuations that deteriorate the resolution. The charge collected in the anode of a triple GEM stack is given by:
\begin{equation}
\label{eq:Q}
Q = N_\mathrm{p}eG_\mathrm{eff1}G_\mathrm{eff2}G_\mathrm{eff3},
\end{equation}
where $N$ is the number of electrons from the primary cloud, $e$ is the electron charge and $G_\mathrm{eff1}$, $G_\mathrm{eff2}$ and $G_\mathrm{eff3}$ are the effective gains of GEMs 1, 2 and 3, as shown in eq.~\ref{eq:geff}.

Assuming the gain of each GEM independent from the others and from the number of primary electrons, one can estimate the detector energy resolution from the charge, applying the error propagation formula to eq.~\ref{eq:Q}:
\begin{equation}
\label{eq:sigmaQ}
\left(\frac{\sigma_Q}{Q}\right)^2 = \left(\frac{\sigma_{N_\mathrm{p}}}{N_\mathrm{p}}\right)^2 + \left(\frac{\sigma_{G_1}}{G_1}\right)^2 + \left(\frac{\sigma_{G_2}}{G_2}\right)^2 + \left(\frac{\sigma_{G_3}}{G_3}\right)^2 .
\end{equation}

The absolute gain of each GEM is the average of the multiplication factor of each independent avalanche $A$:
\begin{equation}
\label{eq:G}
G_{i} = \frac{1}{N_i}\sum_{j=1}^{N_i}A_{ji},\qquad i = 1, 2, 3
\end{equation}
where $N_i$ is the number of electrons available for multiplication in GEM $i$: $N_1 = N_\mathrm{p}$, $N_2 = N_\mathrm{p} G_1$ and $N_3 = N_\mathrm{p} G_1 G_2$. Using the error propagation formula again, this time on eq.~\ref{eq:G}, one gets:
$$
\sigma_{G_i}^{2} = \left(\frac{1}{N_i}\right)^2\sum_{j=1}^{N_i}\sigma_{A_{ji}}^2
$$
\begin{equation}
\sigma_{G_i}^{2} = \frac{1}{N_i} \sigma_{A_i}^2
\end{equation}
Finally, equation \ref{eq:sigmaQ} can be written by:
\begin{equation}
\label{eq:sigmaQfinal}
\left(\frac{\sigma_Q}{Q}\right)^2 = \frac{F}{N_\mathrm{p}} + \frac{1}{N_\mathrm{p}}\left(\frac{\sigma_{A_1}}{G_1}\right)^2 + \frac{1}{N_\mathrm{p} G_1}\left(\frac{\sigma_{A_2}}{G_2}\right)^2 + \frac{1}{N_\mathrm{p} G_1 G_2}\left(\frac{\sigma_{A_3}}{G_3}\right)^2 ,
\end{equation}
where $F = \sigma^2_{N_\mathrm{p}} / N_\mathrm{p}$ is the Fano factor accounting for differences between the observed fluctuations in the number of primary electrons and those predicted by Poisson statistics. Equation~\ref{eq:sigmaQfinal} demonstrates that changing the gain of GEM 1 has much more influence on the energy resolution than changing the gain of any other GEM.

The term 
$$
\left(\frac{\sigma_{A_1}}{G_1}\right)^2 + \frac{1}{ G_1}\left(\frac{\sigma_{A_2}}{G_2}\right)^2 + \frac{1}{G_1 G_2}\left(\frac{\sigma_{A_3}}{G_3}\right)^2 ,
$$
\noindent accounts for the variance in the size of single electron avalanches. It is known that at the electric field regimes of proportional counters, the probability distribution of the number of electrons resulting from single electron avalanche has a deviation from the exponential behavior, being more accurately described by a Polya distribution~\cite{Byr69,Gen73}. This effect has been noticed and studied in GEM based~\cite{Las16} detectors and also in MICROMEGAS~\cite{Zer09}
showing that MPGD have an improved intrinsic energy resolution thanks to their non exponential single electron response.

\begin{figure}[htbp]
\begin{center}
\includegraphics[width=0.96\textwidth]{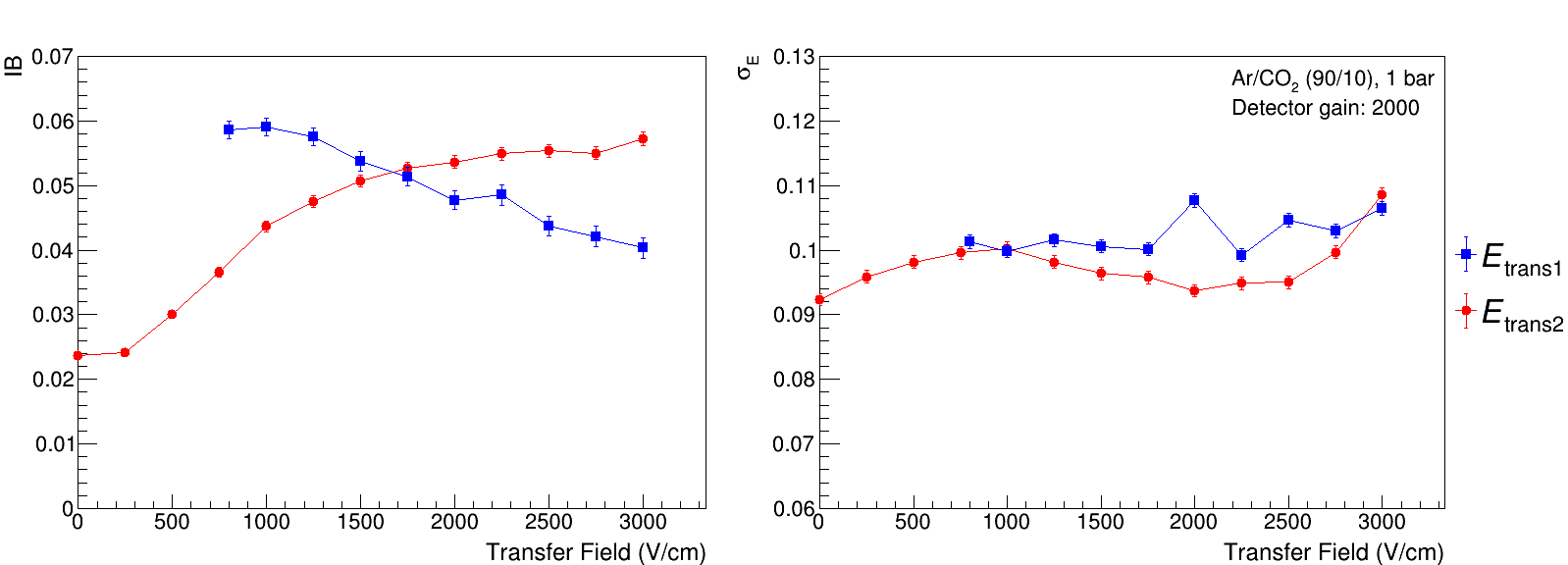}
\caption{Scans of the electric field in both transfer regions ($E_\mathrm{trans1}$: between GEMs LP and S and $E_\mathrm{trans2}$: between GEMs S and SP). The best settings for ion backflow reduction are when $E_\mathrm{trans1}$ is highest and $E_\mathrm{trans2}$ is lowest. The energy resolution does not change considerably. Discussion in the text. }
\label{fig:EtScans}
\end{center}
\end{figure}

Equation~\ref{eq:sigmaQfinal} shows that to improve the energy resolution it is advisable to increase the gain in the first GEM. As mentioned before, this will automatically affect IB.
However there are other parameters that can be tuned which can improve either the energy resolution (or IB), without degrading IB (or energy resolution). In fact, as it can be seen in fig.~\ref{fig:EtScans}, by applying a high field in the first transfer region, between GEMs LP and S and a low field in the second transfer region, between GEMs S and SP, it is possible to minimize the IB, without significantly changing the energy resolution. This has been noticed before in other geometries, and related with the need of applying the highest gain in the last multiplication stage stage~\cite{Ket13,Rat18}, including the ALICE base line configuration~\cite{ALI13}.  The energy resolution is kept constant because the losses in collection efficiency from one GEM are compensated by the gains in the extraction efficiency from the previous GEM, giving origin to a long plateau with constant energy resolution. This is one of the expected consequences of using GEMs with decreasing pitch. The most probable reason for the decrease of the ion backflow when the second transfer field is decreased is, in one hand, the very poor collection of ions in the holes of the LP GEM, due to the high field and, on the other hand, the poor extraction of the ions generated in the SP GEM.

\begin{figure}[htbp]
\begin{center}
\includegraphics[width=0.6\textwidth]{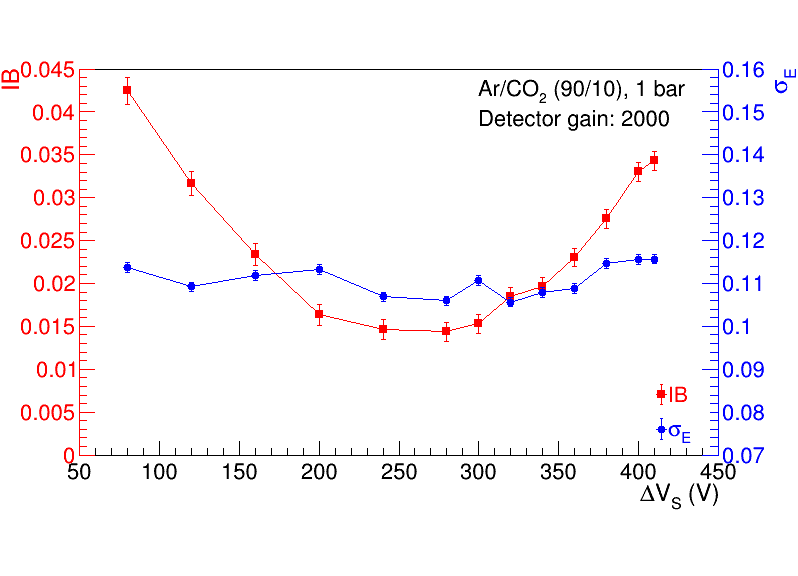}
\caption{Scan of the S GEM. The energy resolution is kept constant, but the ion backflow has an optimal value plateau between 200 and 300\,V. Discussion in the body text.}
\label{fig:Vs}
\end{center}
\end{figure}

Finally the voltage in the S GEM, in the middle of the stack was scanned, keeping the effective gain constant, with a constant \VLP/\VSP ratio. Figure~\ref{fig:Vs} shows again that while the energy resolution remains constant over a wide range of voltages, the ion backflow drops to a plateau of optimal values between 200 and 300\,V. The increase of ion backflow for lower voltages can be explained by the need to increase both \VSP and \VLP. This increases the number of ions generated in the first GEM (LP) which can drift to the cathode. For higher voltages, the collection efficiency of the few ions extracted from the SP GEM is close to unity and these are very well extracted from the S GEM due to the high field in the first transfer region. This can explain the increase of the ion backflow for voltages higher than 300\,V in the S GEM.

\begin{figure}[htbp]
\begin{center}
\includegraphics[width=0.6\textwidth]{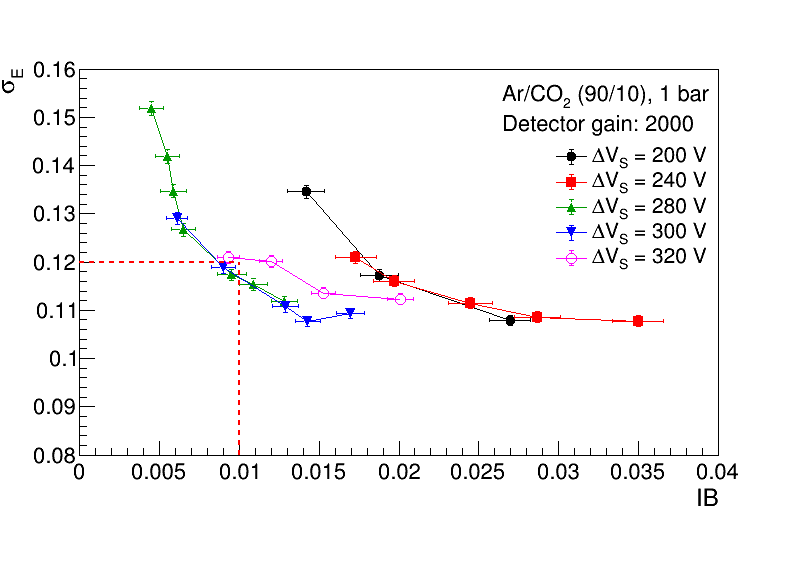}
\caption{\VLP/\VSP scans performed at different voltages applied in the second GEM, after optimizing the transfer fields. It was possible to reach an ion backflow around 1\,\% and an energy resolution around 12\,\%.}
\label{fig:todos}
\end{center}
\end{figure}

By optimizing the transfer fields, several \VLP/\VSP scans were again performed for different voltages in the S GEM, to try to reduce the values of the curve of figure~\ref{fig:VlpVsp}. Figure~\ref{fig:todos} shows the different scans of \VLP/\VSP performed changing the voltage of the S GEM. As expected, as the voltage in the second GEM was close to the values in the plateau, the curves approached the optimal values for IB suppression and energy resolution. It can be seen that for voltages of 280 and 300\,V it was possible to obtain an ion backflow around 1\,\% and an energy resolution of 12\,\%.

\subsection{Simulations}
\label{subsec:simulations}

In conjunction with the experimental measurements presented in this work, Garfield simulation framework~\cite{Vee93} has been used to estimate the energy resolution and ion backflow fraction for the present configuration and have been compared with the available experimental data. The 3D electrostatic field simulation has been carried out using neBEM (nearly exact Boundary Element Method) toolkit~\cite{neBEM01,neBEM02}. Besides neBEM, HEED~\cite{HEED01,HEED02} has been used for primary ionization calculation and Magboltz~\cite{Magboltz01,Magboltz02} for computing drift, diffusion, Townsend and attachment coefficients.

\begin{figure}[htbp]
\begin{center}
\includegraphics[width=0.5\textwidth,trim={0 0 0 0},clip]{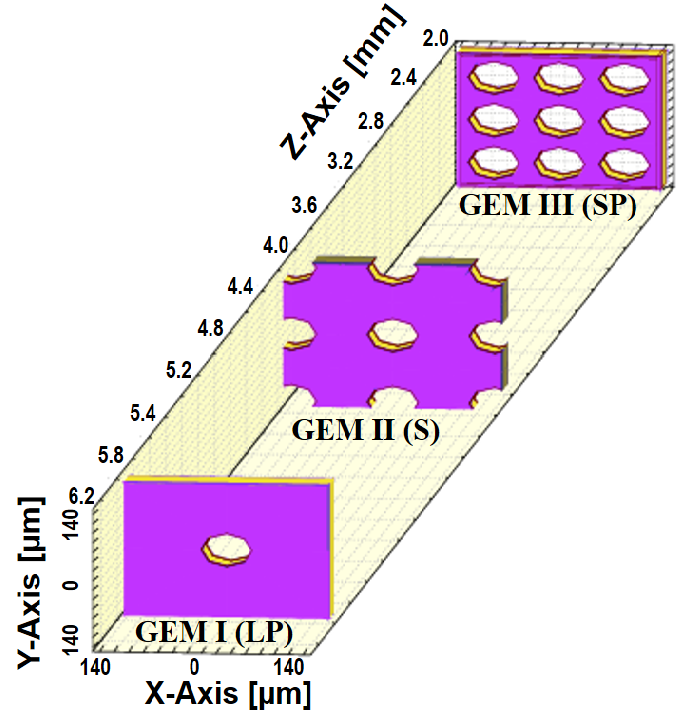}
\caption{Model of the basic GEM cell used in the Garfield simulations.}
\label{fig:sim}
\end{center}
\end{figure}

The model of a basic GEM cell built using Garfield, is shown in fig.~\ref{fig:sim}.
 It represents three GEM foils, having bi-conical shaped holes which are placed in between the drift and the read-out plane. The first GEM after the drift plane has the pitch of 280\,\um, whereas the second and third foils have the pitch of 140 and 90\,\um, respectively. The central holes of the basic units from all the three GEM foils are perfectly aligned. The basic cell structure then has been repeated along both positive and negative X and Y-Axes to represent a real detector. With the help of these models, the field configurations of the detector have been simulated using appropriate voltage settings. These are followed by the simulation of energy resolution and ion backflow fraction.

For estimating electron transmission within GEM detector, electron tracks generated by 5.9\,k\eV photons have been considered in the drift volume. The primary electrons created in the drift region are then made to drift towards the GEM foil and produce avalanche inside the holes. For this calculation Monte Carlo routine has been used. The procedure drifts the primary electrons from their starting position towards the first GEM. Once the field is above the threshold of multiplication, the avalanche starts. At each step, a number of secondary electrons is produced according to the local Townsend and attachment coefficients and the newly produced electrons are traced like the initial electrons. More details of the MC routine used to simulate the avalanche can be found in~\cite{Vee93}. In parallel, the ion drift lines are also traced. The primary ions in the drift region and the ions created in the avalanche have been considered for the estimation of the backflow fraction. 

As mentioned before, the backflow fraction is defined as: 

$$
IB = \frac{N_\mathrm{i}}{N_e},
$$

\noindent where $N_\mathrm{i}$ and $N_e$ are the number of ions collected on the drift plane and the number of electrons collected in the read-out plane, respectively. The 5.9\,k\eV photo-peak of the simulated charge spectrum has been fitted using a Gaussian distribution. The energy resolution has been estimated from the mean and the sigma of this distribution.

\begin{figure}[htbp]
\begin{center}
\includegraphics[width=\textwidth]{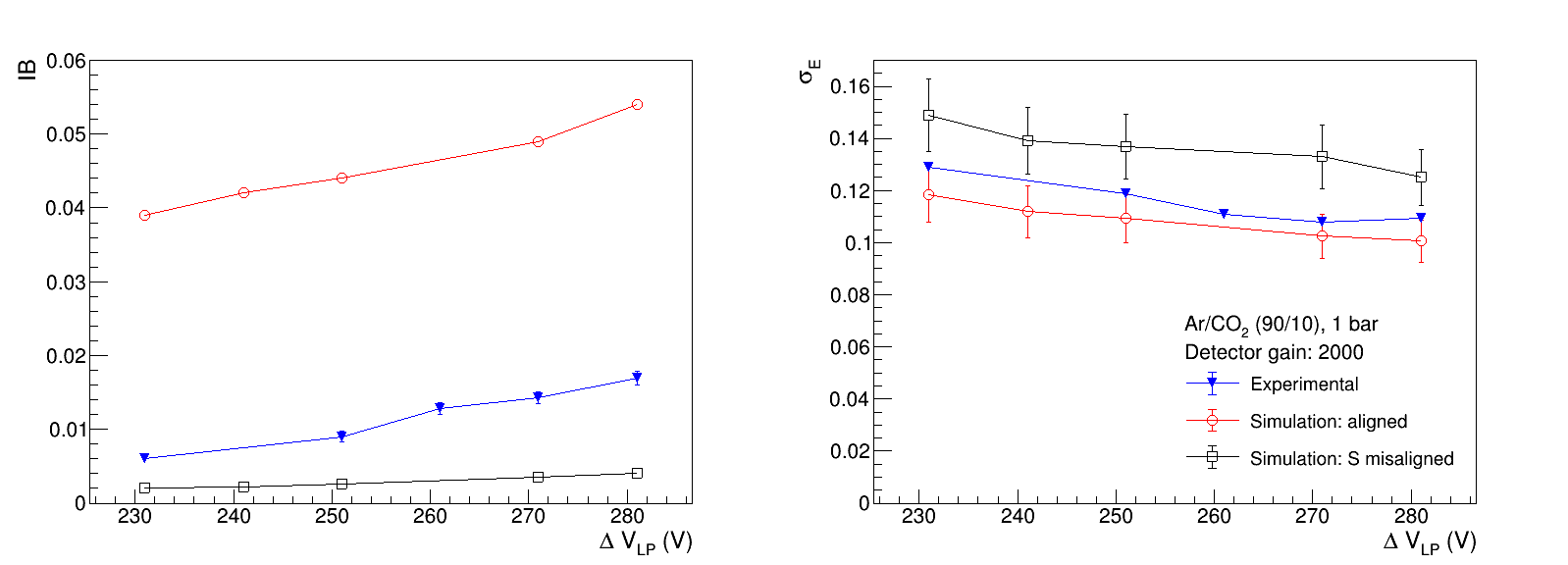}
\caption{Ion backflow fraction and energy resolution obtained with simulation data and compared with the experimental data shown in the previous section. The plot also shows the result obtained when the S GEM is shifted by 70\,\um in the simulation. Discussion in the body text.}
\label{fig:ResIB}
\end{center}
\end{figure}

The variation of the energy resolution and ion backflow fraction with \VLP and \VSP are shown in figure~\ref{fig:ResIB}, for a fixed \VS = 300 V, as in the the experimental data. The large amplification in the first GEM (LP) is the main cause of improvement of the energy resolution, as discussed in section~\ref{subsec:basic}. On the other hand, the increase of the last GEM voltage is also suitable for getting low backflow fraction. Although the simulation data for the energy resolution follow the experimental trend, there is a clear discrepancy for the ion backflow data, where the simulated curve is roughly four times higher than the experimental.

 Reference~\cite{Sau06} shows that the misaligned holes can affect the backflow fraction by one order of magnitude. This was also confirmed in more recent works (for example~\cite{Bal14}) and simulations of a quadruple GEM stack~\cite{Bha17}. The simulations were repeated with the S GEM shifted by 70\,\um in one direction, resulting in misaligned holes. The results obtained are also depicted in the plots and show a 10-fold change in the ion backflow fraction, bringing the simulated curve closer to the experimental results.
This suggests that the main reason for the discrepancy between the experimental and simulated ion backflow results is related to the misalignment of holes in the experimental setup. In fact, besides the impossibility of perfectly align the holes, the first GEM in the stack (LP) is rotated by $90^\circ$ with respect to the other two GEMs due to geometrical constraints of the chamber.

\begin{figure}[htbp]
\begin{center}
\includegraphics[width=0.6\textwidth]{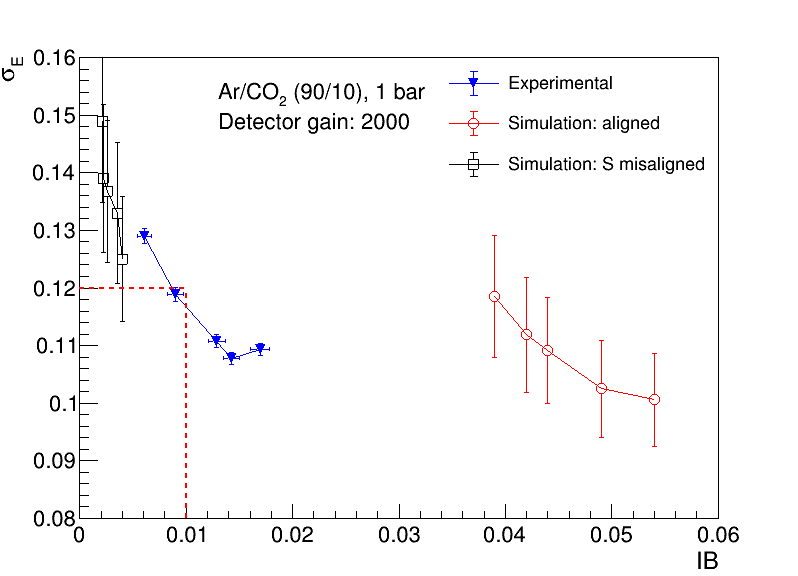}
\caption{Energy resolution and ion backflow fraction obtained with simulation data and compared with the experimental data shown in the previous section. The results obtained when the S GEM is shifted by 70\,\um in the simulation is also shown. Discussion in the body text.}
\label{fig:Compara}
\end{center}
\end{figure}

Figure~\ref{fig:Compara} shows the correlation between the energy resolution and ion backflow fraction. As expected from the previous plots, the results are comparable to the experimental ones when the simulated geometry has the holes of the GEM in the middle misaligned, whereas there is a large discrepancy for the case of aligned holes.

\begin{figure}[htbp]
\begin{center}
\includegraphics[width=0.6\textwidth]{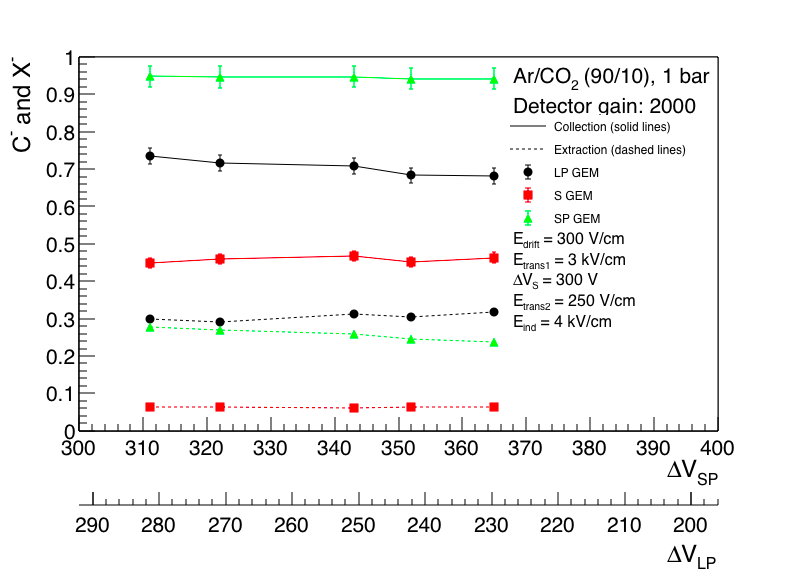}
\caption{Electron collection and extraction efficiencies ($C^{-}$ and $X^{-}$, respectively) for the three GEM foils as the voltages in the LP and SP GEMs are changed. The solid lines are the collection efficiency and the dashed lines are the extraction efficiency.}
\label{fig:EffsV}
\end{center}
\end{figure}

To try to understand what is happening in the detector, the collection and extraction efficiencies ($C^-$ and $X^-$, respectively) as well as the fraction of ions captured in each GEM should be taken into account. In figure~\ref{fig:EffsV} it is possible to see how the electron collection and extraction efficiencies in each GEM vary as the voltages of the LP and SP GEMs change, with the voltage of the S GEM and the transfer fields optimized. At this point of optimization, the values of the efficiencies do not change considerably. This means that the IB is now much more influenced by the gain of each GEM. 

\begin{figure}[htbp]
\begin{center}
\includegraphics[width=0.6\textwidth]{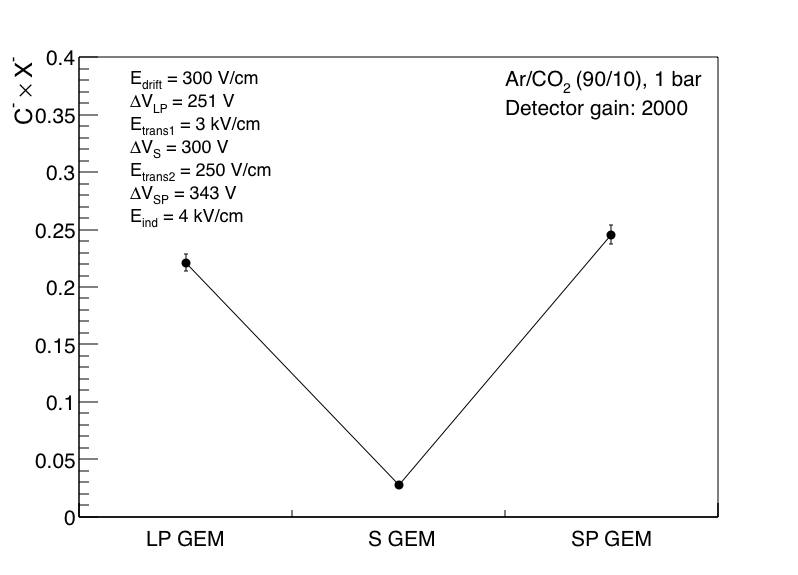}
\caption{The product of the collection efficiency by the extraction efficiency for electrons for the optimal case of 1\,\% IB and 12\,\% $\sigma/E$ in each GEM foil.}
\label{fig:ETransp}
\end{center}
\end{figure}

However, by looking at the data in the perspective of the variation of the product of the electron collection efficiency with the extraction efficiency ($C^{-}\cdot X^{-}$ or $G_\mathrm{abs}/G_\mathrm{eff}$ according to equation~\ref{eq:geff}) as they cross the GEMs, in the optimal voltage settings, it is clear that the already mentioned high voltage settings, where the first transfer field must be high and the second must be low, results in a high electron transmission in the first and third stages and a very low transmission in the intermediate stage, as shown in fig.~\ref{fig:ETransp}. The fraction of ions blocked in each GEM with respect to the total number of ions generated in the avalanches is listed in table~\ref{tab:ions}. Besides showing that only a very small fraction of ions (just below 1\,\%) is not blocked, it is clear that the great majority of ions is blocked already in the SP GEM, thanks to the very low ion extraction field above the SP GEM. The S GEM has a high ion collection field (opposite to what happens with electrons) but the amount of ions to be collected and extracted is already small. The LP GEM blocks the remaining few ions, including those coming from the S GEM.

The very low fraction of ions (with respect to the total number of ions) that reach the cathode suggests that it is the transmission of electrons that must be improved to achieve better IB suppression. Figure~\ref{fig:ETransp} shows there is a clear \emph{bottleneck} in the S GEM. A structure in the place of the S GEM allowing to keep the same fields above and bellow while providing a high transmission of electrons, even at a reduced gain in charge, would be a breakthrough, reducing even more the IB fraction in gaseous detectors, eventually without the need of misaligning the holes.

\begin{table}
\begin{center}
\begin{tabular}{l r}
\hline
GEM 1 (LP) & 0.0450 \\
 GEM 2 (S) & 0.0396 \\
 GEM 3 (SP) & 0.9055\\
 \hline
 Total & 0.9905\\
\hline
\end{tabular}
\caption{The fraction of ions blocked by each GEM foil with respect to the total number of ions produced in the detector.}
\label{tab:ions}
\end{center}
\end{table}

\section{Conclusions}
\label{sec:conclusions}
This proposed geometry for ion backflow suppression using a stack of three GEMs has been able to reduce the flow of ions towards the absorption region to about 1\,\%, while the energy resolution was kept below 12\,\%, operating at a gain of 2000.  

The simple model described in subsection~\ref{subsec:basic} has suggested the advantage of using a GEM with a higher transparency in the last multiplication stage to reduce the ion backflow. Based on that, with the proposed setup we try to take advantage of decreasing hole pitch between the GEMs which would allow to tune the transfer fields at magnitudes promoting a good extraction efficiency while keeping a high collection efficiency in the subsequent stage. This would mean that the second transfer field should be higher than the first. This did not happen and the settings that satisfy the condition for low ion backflow and good energy resolution are those already known and proposed in other studies: one transfer stage with a very high electron extraction efficiency (and consequently low ion collection), followed by a transfer stage with a low field, providing a high electron collection efficiency (and low ion extraction). This is what happens in the transfer zones, suggesting that one of the functions of the S GEM is to provide this transition from high electric field to a smaller one.

Another important feature, also already known and confirmed in this study, is the importance of the misalignment of the holes between different GEMs. The different GEM pitch can promote a larger misalignment, with the result of a smaller ion backflow. This can also be considered as an advantage of this setup. Nevertheless, the fact that the energy resolution was kept constant for a transfer field range larger than usual means that the transfer to a GEM with a smaller pitch is playing an important role. 

One of the constraints of this study was to keep the charge gain of the detector near 2000. This value is relatively small compared to what has been achieved with triple GEM stacks, suggesting that the stability of the system against discharges might not be an issue. However, a dedicated study must be carried out to assess the stability of this system and compare it with the existing solutions. One possible disadvantage can be the fact that the last GEM, with a smaller pitch, has a larger number of holes, where discharges due to defects are more likely to occur. The voltages scanned in the first (LP) and last (SP) GEMs to build the curve from figure~\ref{fig:Compara} were 230 to 280\,V for the first and 310 to 365 for the third GEM foils. These values are well within the range of stable operation of GEMs. For comparison, other large experiments, such as COMPASS and LHCb, have been using triple GEM stacks for several years as high rate trackers, optimized for stability. These GEM based detectors operate at gains well above 2000 and use voltages much higher than the ones reached in this work for the IB and energy resolution optimization (410, 274 and 328 V for GEMs 1, 2 and 3 in COMPASS~\cite{Alt02} and 415 for the three GEMs in LHCb~\cite{Car12}).

Future work will consist on studying if it is possible to divide the transfer regions into a first zone with high electric field increasing the extraction of electrons, followed by a zone with a low field to focus the electrons into the holes of the next GEM. This setup would have the opposite behavior for ions, allowing for a system that would strongly suppress ion backflow, without jeopardizing the energy resolution. The division can be made replacing the S GEM, in the middle, by a dense conductive mesh. Preliminary tests are under way and results will be described in a future publication.

\acknowledgments

H. Natal da Luz acknowledges grant 2016/05282-2, S\~ao Paulo Research Foundation (FAPESP). 
L.A.S.~Filho acknowledges grant 2016/23355-7, S\~ao Paulo Research Foundation (FAPESP).
\noindent This work has partly been performed in the framework of the RD51 Collaboration. We wish to acknowledge the members of the RD51 Collaboration for their help and suggestions.

\noindent We thank our respective Institutions for providing us with the necessary facilities.






\end{document}